\documentclass[aps,prb,reprint,superscriptaddress,longbibliography,showpacs,nobibnotes,nofootinbib]{revtex4-2}

\usepackage[utf8]{inputenc}
\usepackage[T1]{fontenc}
\usepackage[]{lmodern}
\usepackage[]{microtype}
\usepackage{physics}
\usepackage{mathtools}
\usepackage{enumitem}

\usepackage[english]{babel}
\usepackage[]{csquotes}

\makeatletter
\adddialect\l@en\l@english
\makeatother

\usepackage[]{amsmath}
\usepackage[]{amssymb,bbold}
\usepackage[mathscr]{euscript}
\usepackage[percent]{overpic}

\usepackage{siunitx}
\usepackage[]{graphicx}
\usepackage[]{tabularx}
\usepackage[figurename=Figure]{caption}
\usepackage[]{subcaption}

\usepackage{xcolor}

\usepackage{ragged2e}
\DeclareCaptionJustification{plain}{\justifying}
\usepackage[]{hyperref}
\hypersetup{
    colorlinks=true,
    urlcolor=blue,
    citecolor=blue,
    linkcolor=blue
}
\usepackage[capitalize]{cleveref}
\crefname{table}{Tab.}{Tab.}

\DeclareMathOperator{\GEC}{\mathsf{GEC}}

\newcommand{\qmoperator}[1]{\ensuremath{\hat{#1}}}

\newcommand{\ie}{{ i.e.},\ }

\usepackage{comment}

\usepackage{soul}

\usepackage{orcidlink}

\begin{document}
	
\title{Weak ergodicity breaking captured from short-time dynamcis and graph theory}

\author{Heiko Georg Menzler\,\orcidlink{0009-0001-9530-7392}}
\affiliation{Institut für Theoretische Physik, Georg-August-Universität Göttingen, D-37077 Göttingen, Germany}

\author{Rafał Świętek\,\orcidlink{0009-0004-5353-9998}}
\affiliation{Institut für Theoretische Physik, Georg-August-Universität Göttingen, D-37077 Göttingen, Germany}

\author{Mari Carmen Ba\~nuls\,\orcidlink{0000-0001-6419-6610}}
\affiliation{Max-Planck-Institut für Quantenoptik, D-85748 Garching, Germany}
\affiliation{Munich Center for Quantum Science and Technology (MCQST), Schellingstrasse 4, D-80799 München, Germany}

\author{Fabian Heidrich-Meisner\,\orcidlink{0000-0002-3463-1121}}
\affiliation{Institut für Theoretische Physik, Georg-August-Universität Göttingen, D-37077 Göttingen, Germany}

\begin{abstract}
We study the onset of weak ergodicity violations in closed quantum many-body systems
and focus on cases in which they occur through a transition that is controlled by a model parameter, a scenario called a weak ergodicity-breaking transition. This situation is realized in the Rosenzweig-Porter model and the quantum sun model, which we investigate here.
Our analysis is based on representing quantum systems in Fock space and examining the short-time dynamics of density autocorrelation functions. From these, we extract proxies for decay rates
for each individual basis state.
As a first main result, we show that the distribution of such decay rates captures known weak ergodicity-breaking transitions  in both models.
As a second main result, we establish a connection between the decay rates of density autocorrelations and graph-theoretical
measures.
Concretely, we demonstrate that moments of the distribution of  the recently introduced graph-energy centrality  also detect the weak ergodicity transitions, which is inherited from the behavior of the decay rates.
While most numerical tools are limited to small system sizes, the quantities studied in our work can be calculated analytically for large systems of many hundreds of sites and in some cases, even in the thermodynamic limit.
\end{abstract}

\date{\today}

\maketitle

\section{Introduction}
Classifying distinct states of matter is a central goal in condensed matter theory and
quantum many-body physics.
At finite energy densities and temperatures, the essential framework is that of 
spontaneous symmetry breaking~\cite{Binney1992,Goldenfeld1992}. 
This was later extended to quantum phase transitions, where changes in the ground state are driven by quantum rather than thermal fluctuations~\cite{Vojta2003,sachdevbook}.
The second main avenue is the characterization of ground states from the perspective of topology~\cite{Bernevig2013}.
More recently, in the context of closed quantum systems, it has become possible to differentiate between ergodic and nonergodic states of matter at finite energy densities~\cite{Sierant2025}.
Our work is related to this third direction.
In this context, transitions can either be viewed as an analogue of quantum phase transitions in high-energy eigenstates or transitions in the dynamical sense~\cite{Huse2013,Pekker2014,Parameswaran2016}.

To be precise with terminology, we consider a system to be ergodic if it exhibits random-matrix behavior in the sense of the eigenstate thermalization hypothesis (ETH)~\cite{Deutsch1991,Srednicki1994,Rigol2008,DAlessio2016,Deutsch2018,Patil2026}. 
Examples of systems that obey ETH have been studied extensively~\cite{Rigol2009,Rigol2009b,Santos2010,Stenigeweg2013,Khatami2013,Sorg2014,Beugeling2014,Steinigeweg2014,Beugeling2015,Mondaini2016,Mondaini2017,Yoshizawa2018,Jansen2019,Foini2019,Mierzejewski2020,Richter2020,Santos2020,Brenes2020,Brenes2020b,Sugimoto2021,Schoenle2021,Pappalardi2022,Pappalardi2025,Vallini2025}. 
A transition into a nonergodic phase is dubbed an ergodicity-breaking transition (EBT).
The prime candidate system for a nonergodic system is many-body localization~(MBL)~\cite{Luitz2015, Sierant17, Khemani17, Panda2019, Logan19, Colmenarez2019, Corps2021,Abanin2021,Prakash2021}.
In this case, essentially all eigenstates violate ETH.
Strong violations of the ETH in microscopic models also exist  in strong Hilbert-space fragmentation~\cite{Sala2020, Khemani2020, Moudgalya2022c,Lisiecki2025} and integrable interacting systems~\cite{Olshanii2007,Rigol2016}, however, these instances are not robust against perturbations~\cite{Mierzejewski2015,Pandey2020,LeBlond2020,Mallayya2021,LeBlond2021,Bulchandani2022,Surace2023,Kim2024}.
While the existence of MBL as a stable phase of matter is debated~\cite{Sierant2025}, the existence of EBTs is established in 
random matrix models~\cite{Rosenzweig1960,Mirlin1996,Mirlin2000,Evers2000,Kravtsov2015,Bogomolny2018b} and in the quantum sun model (QSM)~\cite{Suntajs2022,Pawlik2024,Pawlik2026}.

Ergodicity can also be weakly broken, when only a subset of eigenstates violates ETH.
Examples include quantum scars~\cite{Turner2018a,Serbyn2021,Desaules2022, Moudgalya2022a,Moudgalya2022b,Channdran2023,Evrard2024}, (weak) Hilbert-space fragmentation~\cite{Rakovszky2020,Jonay2024,Zhang2024,Yang2025,Aditya2025}, and
Fock-space cages~\cite{Tan2025,Benami2025,Jonay2025,Benami2026,Mohapatra2026}.
A particular consequence of such violations are characteristic signatures in autocorrelation functions.
For example, periodic revivals of certain initial states (for quantum scars)~\cite{Turner2018a} or slow relaxation due to constraints (as in fragmented systems) may occur~\cite{Han2026}.
Observing manifestations of such weak ergodicity breaking  is a central theme in ongoing quantum-simulator experiments~\cite{Bernien2017,Kohlert2023,Adler2024,Karch2025,Honda2025}.
In some cases, weak ergodicity violations set in sharply as
a function of model parameters via a weak ergodicity-breaking transition (wEBT)~\cite{Deger2024,DeRoeck2017,Suntajs2022,Kliczkowski2024,Swietek2024,Swietek2025a,Rosenzweig1960,Altland1997,Kravtsov2015,Facoetti2016,Bogomolny2018,DeTomasi2019,Soosten2019,Skvortsov2022,Swietek2025b,Lazarides_2026}.

Recently, it has been demonstrated that the EBT in certain models is preceded by a precursor regime, in which ETH weakly breaks down via a softening of low-frequency fluctuations of matrix elements of local observables.
This scenario is called fading ergodicity and is realized in the Rosenzweig-Porter model (RPM) and the QSM \cite{Kliczkowski2024,Swietek2025a,Swietek2024,Swietek2025b}.
In the RPM, the onset of fading ergodicity coincides with the transition into the well-established fractal phase~\cite{Kravtsov2015}. 
In contrast to EBTs, the onset of the fading-ergodicity regime is not captured by spectrum-averaged random-matrix theory indicators such as the mean gap ratio \cite{Atas2013} but manifests itself in subleading scaling corrections encoded in their full distributions~\cite{Swietek2025b}.
However, resolving scaling corrections on mid-spectrum properties using finite-size exact diagonalization (ED) is
a challenging and often unfeasible task, see, e.g.,~\cite{Sierant2025}, even with pure-state algorithms~\cite{Pietracaprina2018,Sierant2020b,Pintar2026}.
\begin{figure}[b]
    \centering
    \includegraphics{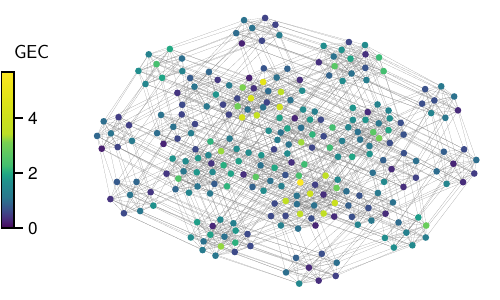}
    \caption{
        Fock-space graph of one instance of the QSM ($L=5$, $N=3$, $\alpha=0.8$) with the $\GEC$ values computed for each node corresponding to basis states (see the color~bar).
    }
    \label{fig:introfigure}
\end{figure}

In this work, we are interested in \textit{sudden} weak ETH violations
as they appear in the RPM and the QSM via a wEBT. We choose these two models since the existence of both wEBTs and EBTs is established~\cite{Suntajs2022,Kliczkowski2024,Swietek2024,Swietek2025a,Rosenzweig1960,Kravtsov2015,DeTomasi2019,Skvortsov2022,Venturelli2023}.
As a main result, we show that the short-time expansion of density autocorrelation functions carries information about wEBTs at infinite temperature in these models.
Concretely, we extract approximated decay rates, which we denote as $1/\tau$, for individual Fock-space states and study their distribution over the entire many-body basis and Hamiltonian realizations. 
The calculation of the $1/\tau$ does not require the diagonalization of a  Hamiltonian.
The typical value of $1/\tau$ shows a sharp change in behavior at the wEBT.
In the thermodynamic limit, these features translate to a sharp transition at the wEBT in both models.
Concretely, in the RPM, the typical value of the $1/\tau$ jumps discontinuously at the onset of fading ergodicity in the thermodynamic limit,
whereas for the QSM, the typical value of the $1/\tau$ remains continuous, but is not differentiable at the wEBT.

Additionally, we investigate the behavior of the recently introduced \emph{graph-energy centrality} ($\GEC$)~\cite{Menzler2025}, 
inspired from know centrality measures \cite{Das2018}.
This object was originally introduced to measure the competition of diagonal and off-diagonal terms in a given basis that is used to set up the Fock-space graph.  A typical graph is illustrated in \cref{fig:introfigure}.  Through similarities in changes of distributions and abrupt changes in their moments, we show that the $\GEC$ can be equally well used as a measure for predicting wEBTs in the RPM and the QSM, and more generally, for fading ergodicity. Furthermore, we argue that the GEC inherits its behavior from the $1/\tau$  and thus we provide a physical interpretation of this graph-theoretical measure. 

Most importantly from the methodological point of view, we demonstrate that leading moments of the distribution of $1/\tau$ values and of the GEC can be computed analytically for large systems of many hundreds of sites and sometimes even in the thermodynamic limit.

The remainder of this work is structured as follows.
In \cref{sec:preliminaries}, we introduce the main quantities studied here.
We start by introducing density autocorrelation functions in initial Fock-states and define the decay rates $1/\tau$ from their short-time expansion in \cref{sec:preliminaries:tau}.
In \cref{sec:preliminaries:gec}, we define the $\GEC$ for individual Fock states, we define the models of interest in \cref{sec:preliminaries:models}, while in \cref{sec:preliminaries:calculations}, we provide details on the computation of both the coefficients $1/\tau$ and the $\GEC$.
Next, we study the RPM in \cref{sec:rpm} using the measures defined in \cref{sec:preliminaries}.
In \cref{sec:qsm}, we repeat the same analysis for the QSM, where we first study diagonal matrix elements of a spin-spin correlator [\cref{sec:qsm:eth}], 
followed by identifying the wEBT [\cref{sec:qsm:webt}].
We conclude in \cref{sec:conclusions}.
In the \cref{app:chi2,app:gec,app:gec:rpm,app:tau:qsm},we provide details on the analytical derivation of the typical value of $1/\tau$ for both models and the typical value of the $\GEC$ for the QSM.
We also show the mean and the variance of the $\GEC$ for the QSM in \cref{app:qsm:gec:variance}.

\section{Short-time dynamics and GEC}\label{sec:preliminaries}
\subsection{Short-time expansion of autocorrelation functions}\label{sec:preliminaries:tau}
Without loss of generality, we assume that the Fock states $\ket{i}$ considered in the following are basis states from the joint eigenbasis of local density operators $\hat{n}_\ell$.
Then, we define density autocorrelation functions in the real-space occupation basis,
\begin{equation}\label{eq:autocorrelation}
    C_{i,\ell}(t)=\mel{i}{\hat n_\ell(t)\hat n_\ell}{i}\,.
\end{equation}
In the following, we will always consider a fixed site $\ell$ but  different initial states $\ket{i}$.

The short-time expansion of the autocorrelation function in \cref{eq:autocorrelation} gives (henceforth, we set $\hbar=1$)
\begin{equation}\label{eq:autocorrelation:short_t:full}
    C_{i,\ell}(t)=\mel{i}{\hat n_\ell^2}{i}
    -\frac{t^2}{2}\mel{i}{[\hat H,[\hat H,\hat n_\ell]]\hat n_\ell}{i}
    +O(t^4).
\end{equation}
Using that for a local density operator $\hat n_\ell^2=\hat n_\ell$ and using $\hat n_\ell\ket{i}=n_\ell^{(i)}\ket{i}$, we find that the leading decay of $C_{i,\ell}(t)$ can equivalently be written as
\begin{equation}\label{eq:autocorrelation:short_t}
    C_{i,\ell}(t)=n_\ell^{(i)}-t^2/\tau^2_{i,\ell}+O(t^4),
\end{equation}
with
\begin{equation}\label{eq:tau:def}
    \frac{1}{\tau_{i,\ell}^2}
    =-n_\ell^{(i)}\mel{i}{[\hat H,\hat n_\ell]^2}{i}
    =n_\ell^{(i)}\left[\mel{i}{\hat H^2}{i}
    -\mel{i}{\hat H\hat n_\ell\hat H}{i}\right]\,.
\end{equation}
Ref.~\cite{Kim2015} shows that the Frobenius norm of the commutator in \cref{eq:tau:def} provides an upper bound for relaxation rates at high temperatures.
We propose that the $1/\tau_i$ from \cref{eq:tau:def} can  be considered as a short-time decay rate for individual initial states.
We note that similar approaches to study slow dynamics in the short-time expansion have been the subject of recent studies~\cite{Kim2015, Pawlowski2025}.

Since we choose a local density operator for the autocorrelation functions $C_{i,\ell}(t)$, for all initial states that are unoccupied at the given site, $C_{i,\ell}(t)$ vanishes identically.
Such initial states are not considered in the distribution of $1/\tau$.

\subsection{Graph-energy centrality}\label{sec:preliminaries:gec}
The crucial step in defining $\GEC$ is to represent a given Hamiltonian $\qmoperator H$ as a weighted graph in Fock space.
Choosing a  basis, nodes correspond to basis states $\ket{i}$ and the weight of the edge between two nodes $\ket{i}$ and $\ket{j}$ corresponds to the matrix element $\mel{i}{\qmoperator H}{j}$.
We allow for both positive and negative weights and we interpret diagonal matrix elements $\mel{i}{\qmoperator H}{i}$ as edges connecting a state $\ket{i}$ to itself.
In order to make a connection to a given physics problem, it will be important to choose an appropriate basis.

The $\GEC$ for a basis state $\ket{i}$ is defined as~\cite{Menzler2025} 
\begin{align}\label{eq:gec:def}
    \GEC_i = \frac{\Tr(\qmoperator H^2) - \Tr([\qmoperator H\setminus \ket{i}]^2)}{\Tr(\qmoperator H^2)/D}\,,
\end{align}
where $\qmoperator H\setminus \ket{i} \coloneq (\mathbb{1} - \ketbra{i}{i})\qmoperator H(\mathbb{1}-\ketbra{i}{i})$ projects the state~$\ket{i}$ out of the Hilbert space of dimension $D$.
For calculating the $1/\tau$ and $\GEC$, we scale the Hamiltonian as $\qmoperator H \to (\qmoperator H - \mu\mathbb{1})/\sigma$, where $\mu = \Tr(\qmoperator H)/D$ and
\begin{equation}\label{eq:sigma}
    \sigma^2 = \Tr(\qmoperator H^2)/D-\mu^2\,.
\end{equation}
This results in a traceless Hamiltonian with eigenvalues of unit variance.
Note that the definition in \cref{eq:gec:def} differs from the one introduced in~\cite{Menzler2025} by a factor of $D$.
The reason for this change lies in the fact that the $\GEC$ as defined in Ref.~\cite{Menzler2025} decays with the Hilbert-space dimension as $1/D$.
On the one hand, the numerator of the $\GEC$ in \cref{eq:gec:def} can be rewritten as the difference of matrix elements $\Tr(\qmoperator H^2) - \Tr([\qmoperator H\setminus \ket{i}]^2)=2\mel{i}{\hat{H}^2}{i}-\mel{i}{\hat{H}}{i}^2$.
On the other hand, the denominator represents a sum of such matrix elements $\Tr(\qmoperator H^2)=\sum_i\mel{i}{\hat{H}^2}{i}$ and is therefore always $D$ times larger than the numerator.
The additional factor of $D$ removes this trivial scaling.

The rescaling of the Hamiltonian with the variance in \cref{eq:sigma} allows us to compare the decay rates $1/\tau$ across different system sizes and model parameters.
In other words, we express the decay rates in units of $\sigma$, which is the relevant energy scale of the system.
In certain parameter regimes (such as the ergodic phase for the RPM), the bandwidth scales exponentially with system size, while in other phases, it does not (for example, in the localized phase of the RPM).
By expressing the timescales in units of $\sigma$, we remove this scaling and are able to compare the relevant timescales across the entire phase diagram of the model.

\subsection{Models}\label{sec:preliminaries:models}
\subsubsection{Rosenzweig-Porter model}
The RPM~\cite{Rosenzweig1960,Altland1997,Kravtsov2015,Facoetti2016,Bogomolny2018,DeTomasi2019,Soosten2019,Skvortsov2022} is described by the Hamiltonian 
\begin{equation}\label{eq:rpm:ham}
    {\qmoperator H_\mathrm{RPM} = \qmoperator H_0 + D^{-\gamma/2} \qmoperator M}\,,
\end{equation}
consisting of a $D\times D$ dimensional diagonal matrix $\qmoperator H_0$ with entries drawn from a standard normal distribution and an off-diagonal term $\qmoperator M$. The latter is a random matrix sampled from the Gaussian orthogonal ensemble (GOE) $\qmoperator M = (\qmoperator B + \qmoperator B^\dagger)/\sqrt{2}$, where all entries of the matrix $\qmoperator B$ are drawn from a standard normal distribution.
Notice that our definition of the RPM includes the diagonal in the GOE matrix, following~\cite{Kunz1998,Barney2023,Venturelli2023,Cadez2024,Kutlin2025}.
The RPM features two transitions~\cite{Kravtsov2015}.
First, at $\gamma=1$, the RPM changes from a phase obeying random matrix behavior with extended ergodic eigenstates ($\gamma<1$) into a regime ($1 < \gamma < 2$) exhibiting fading ergodicity~\cite{Swietek2025b} and where eigenstate properties show fractal scaling behavior ~\cite{DeLuca2014,Kravtsov2015,Truong2016,Bogomolny2018,DeTomasi2019}.
This regime is also referred to as   non-ergodic extended phase in the literature~\cite{Kravtsov2015}.
Second, a transition from the fractal regime to a fully localized regime occurs at $\gamma = 2$.
ETH is valid in the strong sense for $\gamma < 1$ and is weakly violated for $1 < \gamma < 2$~\cite{DeTomasi2020,Skvortsov2022,Venturelli2023,Buijsman2024}.

Similar to previous approaches~\cite{Swietek2025b} we embed the RPM Hamiltonian in the Hilbert space of $L$ sites.
We construct a diagonal matrix $\hat{H}_0$, where each entry is labeled by an index $i=0,...,2^L-1$.
The corresponding Fock states $\ket{i}$ are  the bit representations of the integer $i$ with a fixed length $L$.
In this way the notion of a lattice site is obtained by the position in the bit representation of the integer corresponding to a particular basis state.
It should be noted that, since there is no natural notion of sites in the RPM, the choice of ordering the sites is not unique and the representation does not constrain the model to a certain spatial dimension. 
Due to the non-local character of the model, all such representations are equivalent.
A recent work discussed different approaches of labeling many-body basis states in the example of a different random matrix ensemble~\cite{Buijsman2026}.

Following this representation, a basis state $\ket{i}$ can be expressed by
$\ket{i}\equiv\ket{i_1,\cdots i_L}=\ket{i_1}\otimes\cdots\otimes\ket{i_L}$, where each $\ket{i_\ell}\in\mathcal{H}_\ell\equiv\{\ket{0}_\ell,\ket{1}_\ell\}$.
The entire Hilbert space is then a tensor product $\mathcal{H}=\otimes_{\ell=1}^L\mathcal{H}_\ell$.
We define a local density operator $\hat{n}_\ell$ that takes value $n_\ell=1$, when the bit at position $\ell$ is set to one, and $n_\ell=0$, when that bit is set to zero.
By construction the density operators on different sites commute and are orthogonal.
These operators $\hat n_\ell $ are used to compute the density autocorrelation functions.

\subsubsection{Quantum sun model}
Now we consider a model that, in contrast to the RPM, is composed of few-body operators.
The Hamiltonian of the QSM describes an ergodic bubble of $N$ spin-$1/2$ objects coupled to $L$ localized spins~\cite{Suntajs2022,Swietek2024} and is expressed as
\begin{equation}\label{eq:qsm}
    {\qmoperator H_\mathrm{QS} = \qmoperator R + g_0 \sum_{\ell=1}^L \alpha^{u_\ell} \qmoperator S^x_{k(\ell)} \qmoperator S^x_{\ell} + \sum_{\ell=1}^L h_\ell \qmoperator S^z_\ell}\,.
\end{equation}
$\qmoperator R$ is a normalized GOE matrix ($\qmoperator R = \qmoperator M/\sqrt{2^N+1}$, where $\qmoperator M$ is of size $2^N \times 2^N$) representing the ergodic bubble, which is coupled to the $L$ localized spins through a coupling parameter $g_0$.
The total number of spins is $L_{\rm tot}=L+N$, such that the Hilbert space grows as $D=2^{L+N}$.
We label the spins inside the bubble with non-positive numbers $k(\ell)=-N+1,\ldots,0$, while the localized spins outside are labeled with $\ell=1,\ldots,L$.
$\hat S_{\ell}^\kappa$ and $\hat S_{k(\ell)}^\kappa$ are the components of spin-1/2 operators acting on a localized spin at site $\ell$ and a spin in the bubble at site $k(\ell)$, respectively, with $\kappa=x,y,z$.
Each localized spin at site $\ell$ couples to a randomly selected spin at site $k(\ell)$ inside the bubble with a coupling strength $\alpha^{u_\ell}$, where $u_{\ell}$ is uniformly sampled from the interval $[\ell-1-\zeta, \ell-1+\zeta]$ with $\zeta=0.2$, except for the first site, for which $u_1=0$.
Furthermore, spins outside the ergodic bubble are subject to a random magnetic field $h_\ell$ sampled uniformly from an interval $[h-W,h+W]$, where we choose $h=1$, $W=0.5$ and $g_0=1$.

For the QSM, an EBT is analytically predicted to occur at $\alpha=1/\sqrt{2}$~\cite{DeRoeck2017,Swietek2025a}, with Poisson level statistics for $\alpha\ll1/\sqrt{2}$~\cite{DeRoeck2025}.
The ETH is valid in a strong sense at $\alpha = 1$,
while the entire regime $1/\sqrt{2} < \alpha < 1$ is described by fading ergodicity, where relaxation times become exponentially large, yet smaller than the Heisenberg time $t_H=2\pi/\Delta$, where $\Delta$ is the mean level spacing~\cite{Kliczkowski2024}.
This implies a wEBT at $\alpha = 1$.
Since wEBTs are the main subject of this work, we study the QSM for interaction strengths $0.6 < \alpha < 1.3$.

\subsection{Calculating the $1/\tau$ and $\GEC$}\label{sec:preliminaries:calculations}

The evaluation of $\GEC$ can be simplified to ${\GEC_i \propto 2\mel{i}{\qmoperator A^2}{i} + \mel{i}{\qmoperator H}{i}^2}$, where $\qmoperator A$ is the off-diagonal part of $\qmoperator H$ in the chosen basis.
For the quantum East model~\cite{Menzler2025}, these terms are only a function of the number of particles in the basis states, which allows an efficient exact calculation of the \textit{entire} $\GEC$ distribution on large system sizes beyond the reach of exact diagonalization.

\begin{figure}[t]
    \centering
    \includegraphics[width=\columnwidth]{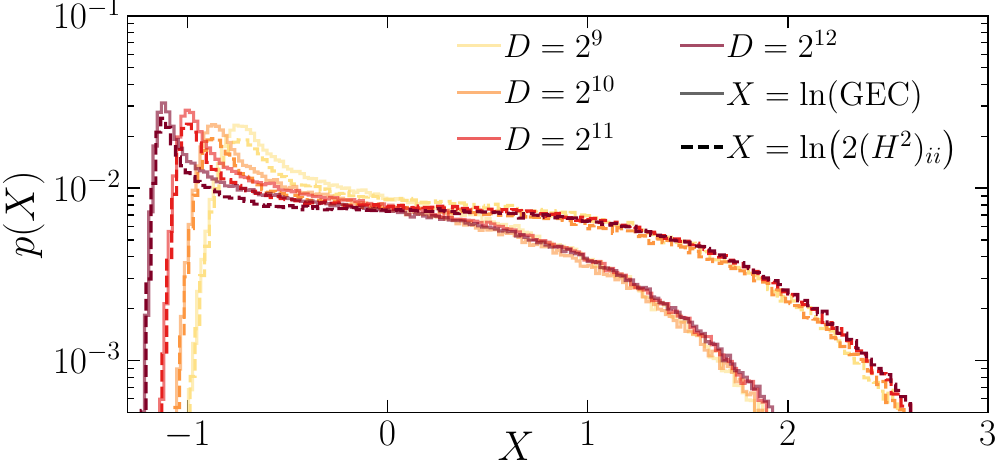}
    \caption{Rosenzweig-Porter model.
        Distribution of the $\GEC$ [solid line] and the diagonal matrix elements $(H^2)_{ii}$ [dashed line] that enter the $\GEC$ for a fixed value of $\gamma=1.2$ in the fading-ergodicity regime for different Hilbert-space dimensions.
    }
    \label{fig:RP_tau:a}
\end{figure}
The models considered in this work [see \cref{sec:preliminaries:models}] are random Hamiltonians sampled from a certain ensemble.
Specifically, for the RPM, we sample the matrix elements in \cref{eq:rpm:ham} and use their distributions to calculate the typical values of the $\GEC$ and $1/\tau$.
For the QSM model, we draw random numbers for each local term in the Hamiltonian in \cref{eq:qsm}.
In this case, we are able to evaluate the typical value of $1/\tau$ by calculating the commutator of the Hamiltonian with a local density operator.

Carrying out the ensemble average of the decay rates $1/\tau$ and the $\GEC$ requires an approximation.
In these cases, we replace the standard deviation of the energies $\sigma$ in \cref{eq:sigma} with the ensemble averaged variances $\sigma\to\sqrt{\overline{\sigma^2}}$, where $\overline{\vphantom{sum}\ldots}$ is an average over disorder realizations (see the discussion in  \cref{app:gec}).
With this approach, we can compute the mean, variance and the typical value of the distributions $p(1/\tau)$ and $p(\GEC)$  for the RPM analytically for large system sizes in the thermodynamic limit (for a discussion, see \cref{app:gec:rpm}).
Specifically, we utilize the distributions of the matrix elements of the RPM to calculate the moments of both quantities.
For the QSM model, we are able to evaluate the commutator in \cref{eq:tau:def} exactly and are left in finding the ensemble averaged width of the spectrum $\overline{\sigma^2}$ (see \cref{app:tau:qsm} for details).
The typical value of the $\GEC$ distribution for the QSM is computed numerically.
\begin{figure}[b]
    \centering
    {\phantomsubcaption\label{fig:RP_tau:b}}
    {\phantomsubcaption\label{fig:RP_tau:c}}
    \includegraphics[width=\columnwidth]{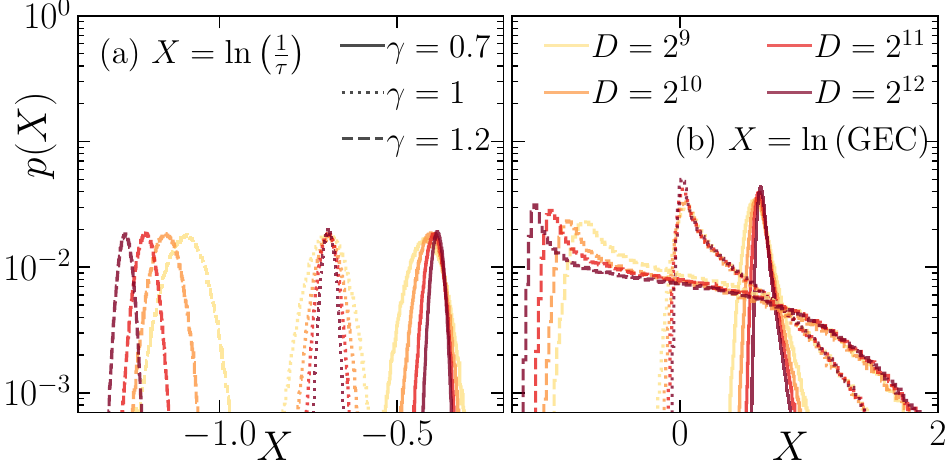}
    \caption{Rosenzweig-Porter model.
        \subref{fig:RP_tau:b} Distribution of coefficients $1/\tau$ in the short-time expansion of the density autocorrelation functions [see \cref{eq:tau:def}], and
        \subref{fig:RP_tau:c} distribution of the $\GEC$ values for different values of $\gamma$ and different matrix dimensions [see \cref{eq:gec:def}]. 
    }
    \label{fig:RP_tau}
\end{figure}

Both $\GEC_i$ and $1/\tau_i$ contain the matrix element $\mel{i}{\hat H^2}{i}$, which measures the total weight of matrix elements connecting $\ket{i}$ to other Fock-space configurations.
More importantly, the term $\mel{i}{\hat H^2}{i}$ controls the decay rate at short times.
This can be understood from the fact that these matrix elements are semi-positive definite and appear in the only term in \cref{eq:tau:def} that can lead to a decay of the autocorrelation function $C_{i,\ell}(t)$.
Additionally, the features observed in the moments of the $\GEC$ can also be  found in the moments of $\mel{i}{\hat H^2}{i}\equiv (H^2)_{ii}$.
We present these similarities in \cref{fig:RP_tau:a} and discuss them in detail in \cref{sec:rpm}.
This connection allows us to relate the $\GEC$ to the decay rates $1/\tau$.
We thus interpret the $\GEC$ as a measure of early-time dynamics in  Fock space, where small $\GEC$ values indicate an increase of  relaxation times (corresponding to small $1/\tau$) of autocorrelation functions.

\section{Rosenzweig-Porter model}\label{sec:rpm}
In this section we study the distributions and typical values of $1/\tau$ and the $\GEC$ for the RPM in \cref{eq:rpm:ham}.
We show that both quantities capture the wEBT at $\gamma=1$.
Without loss of generality, we consider the density operator at site $\ell=L/2$ in the short-time dynamics.
Since the RPM is an all-to-all connected matrix, choosing any site leads to identical results after ensemble averaging, where each site corresponds to a bit in the binary representation of the $\hat H_0$-basis indices, see \cref{sec:preliminaries:models}.
Correspondingly, we use the same basis to compute the GEC.

\begin{figure}[t]
    \centering
    {\phantomsubcaption\label{fig:RP_var_gec_tau:a}}
    {\phantomsubcaption\label{fig:RP_var_gec_tau:b}}
    \includegraphics[width=\columnwidth]{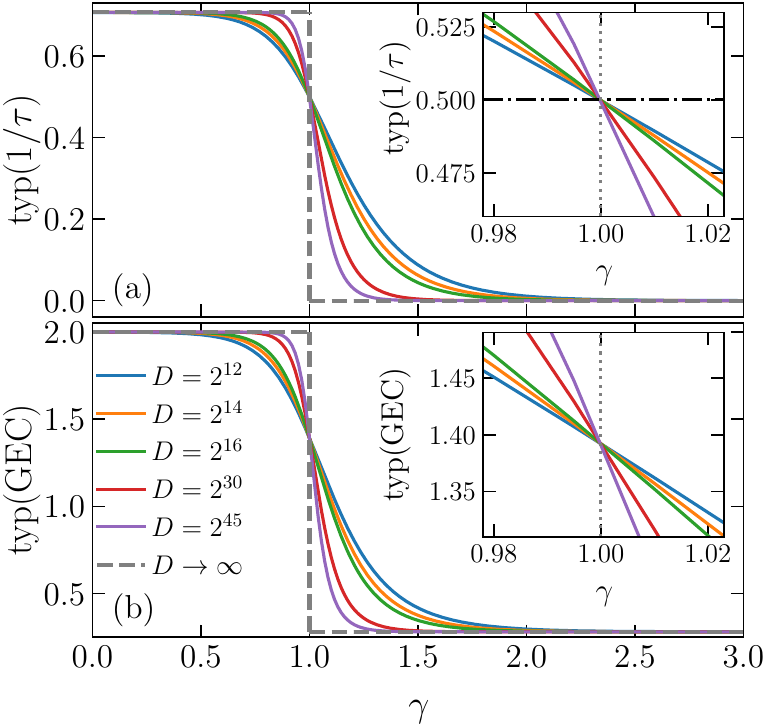}
    \caption{
        \subref{fig:RP_var_gec_tau:a} Rosenzweig-Porter model. Typical value of the coefficients $1/\tau$ in the short-time expansion of the density autocorrelation functions and \subref{fig:RP_var_gec_tau:b} typical values of the $\GEC$ distribution for different values of $\gamma$ and different system sizes.
        The insets in both panels show the vicinity of the crossing point of the curves for different matrix dimensions.
    }
    \label{fig:RP_var_gec_tau}
\end{figure}

In \cref{fig:RP_tau:b}, we display the distributions of $1/\tau$  for $\gamma<1$ and $\gamma \geq 1$, for various system sizes.
The distributions of $1/\tau$ can be well approximated by a 
$\chi^2_{D/2}$-distribution that becomes Gaussian in the $D\to\infty$ limit, see \cref{app:gec:rpm}.
Since the RPM couples every basis state to all others, the distribution of $1/\tau$ remains similar across different initial product states, indicating the absence of strongly state-dependent relaxation rates.
The progressive shift of the distributions toward smaller values of $1/\tau$ as $\gamma$ increases signals an increasingly slow decay of the density autocorrelation functions. 
This systematic suppression provides a dynamical signature of weak ergodicity breaking.

The changes in the distributions of $1/\tau$ are reflected in corresponding features of the distribution of the $\GEC$, shown in \cref{fig:RP_tau:c}. Namely, maximum of the $\GEC$ distribution moves towards smaller values as $\gamma$ increases.
In the ETH and in the fading-ergodicity regime the distributions of the $\GEC$ are distinctly different from each other.
This observation can be understood from
limiting cases of $\GEC$ distributions at $\gamma=0$ and $\gamma\gg1$.
For $\gamma=0$, the distribution is a narrow Gaussian, since the $\GEC$ can be described by a sum of $D$ squared Gaussian random variables,
corresponding to all entries in a row of the GOE matrix $\qmoperator M$ plus one element of $\qmoperator H_0$.
This distribution can be asymptotically described by a $\chi^2_{D}$-distribution with $D$ degrees of freedom.
In the limit $D\to\infty$, the distribution becomes Gaussian and narrow, as a consequence of the central limit theorem.
For $\gamma \gg 1$, the diagonal terms dominate.
Unlike the $\gamma=0$ case, there is only a single contributing squared Gaussian random variable, which makes the expected distribution a single-degree-of-freedom $\chi^2_1$-distribution.

Next, we argue that the distributions of GEC inherit the behavior from the one of the $1/\tau$. To that end, we study the similarities of the distributions of $\GEC$  and $(H^2)_{ii}$ values in \cref{fig:RP_tau:a}.
We choose a fixed value of $\gamma=1.2$, but we observe the same common features in those distributions across all values of $\gamma$ studied in this work.
Since the definition of the $\GEC$ contains two times the diagonal element $(H^2)_{ii}$ we compare the distribution of the $\GEC$ to the distribution of $2(H^2)_{ii}$.

By inspecting \cref{fig:RP_tau:a}, we notice that both distributions  possess heavy tails for large values of their components, with the $\GEC$ admitting smaller values due to the additional contributions from $H_{ii}^2$.
As a result, both quantities have different means and variances.
Nevertheless, their typical values, which are of main interest in this work, are identical and follow the same system size trends, as shown in \cref{fig:RP_var_gec_tau}.
Therefore, the  features  relevant for the wEBT are shared by the distributions of both $\GEC$  and $(H^2)_{ii}$.

Having clarified the connection between $1/\tau$, $(H^2)_{ii}$, and 
$\GEC$, we address the behavior of $p(1/\tau)$ and $p(1/\GEC)$ in the vicinity of the wEBT.
The differences in their respective distributions translate into critical behavior of the moments of $1/\tau$ and the $\GEC$ at the transition into the fading-ergodicity regime.
Using analytical arguments, we obtain the typical value $\mathrm{typ}(1/\tau)$ and the typical value 
$\mathrm{typ}(\GEC)$ for arbitrary $D$, see \cref{app:gec:rpm} for the derivation.
\Cref{fig:RP_var_gec_tau:a,fig:RP_var_gec_tau:b} show  the results for $\mathrm{typ}(1/\tau)$ and $\mathrm{typ}(\GEC)$ as a function of $\gamma$, respectively.
In the thermodynamic limit ($D\to\infty$), $\mathrm{typ}(1/\tau)$ becomes a step function that jumps from $\lim_{D\to\infty}\mathrm{typ}(1/\tau)=1/\sqrt{2}$ for $\gamma<1$ to $\lim_{D\to\infty}\mathrm{typ}(1/\tau)=0$ for $\gamma>1$, see \cref{fig:RP_var_gec_tau:a}.
For finite systems, there is a crossing point of $\mathrm{typ}(1/\tau)$ at $\gamma=1$, where $\mathrm{typ}(1/\tau)=1/2+O(D^{-1})$.
The inset in \cref{fig:RP_var_gec_tau:a} shows the vicinity of the crossing point,
where curves for different values of $D$ perfectly intersect.
In \cref{fig:RP_var_gec_tau:b}, we observe identical characteristics for  $\mathrm{typ}(\GEC)$.
The distribution of $\GEC$ becomes very narrow with a finite typical value $\lim_{D\to\infty}\mathrm{typ}(\GEC)=2$ at $\gamma<1$, while for $\gamma>1$ the variance stays finite and the typical value flows to $\lim_{D\to\infty}\mathrm{typ}(\GEC)=e^{-c_E}/2$, with $c_E\approx0.57721$ being the Euler's constant.
The inset of \cref{fig:RP_var_gec_tau:b} shows a closed-in view of the crossing point of $\mathrm{typ}(\GEC)$ at $\gamma=1$.

The significance of the $1/\tau$ values extends beyond the short-time expansion of density autocorrelation function in \cref{eq:autocorrelation:short_t}.
In the RPM (and also in the QSM) there are no conventional hydrodynamic modes that generate power-law decay of autocorrelation functions, as a consequence of the all-to-all coupling in the model.
In other words, there are no additional dynamical features emerging beyond the short-time decay until the system relaxes.
This distinction will be particularly important when considering local Hamiltonians with additional conservation laws.

\begin{figure}[t]
    \centering
    \includegraphics[width=0.99\columnwidth]{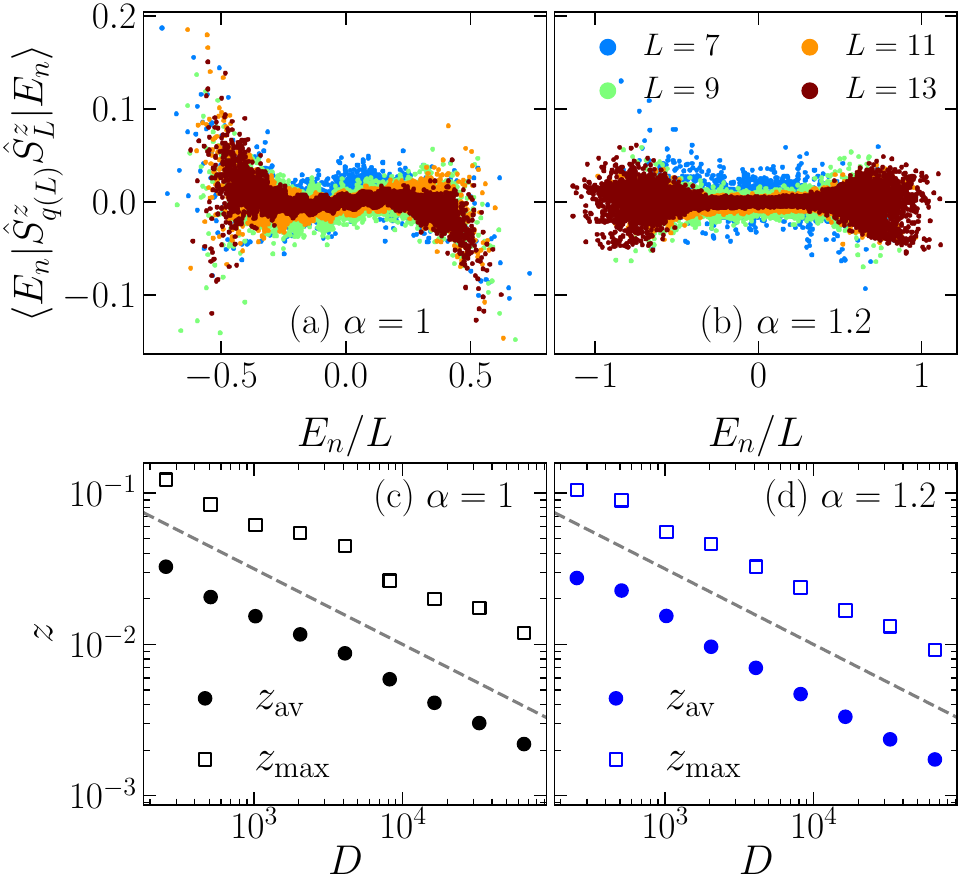}
    {\phantomsubcaption\label{app:fig:qsm:alfa=1}}
    {\phantomsubcaption\label{app:fig:qsm:alfa=1.2}}
    {\phantomsubcaption\label{app:fig:qsm:fluct:alfa=1}}
    {\phantomsubcaption\label{app:fig:qsm:fluct:alfa=1.2}}
    \caption{Quantum sun model. Eigenstate expectation values for the observable $\hat{S}^z_{q(L)}\hat{S}^z_L$ in the QSM for \subref{app:fig:qsm:alfa=1} $\alpha=1$ and \subref{app:fig:qsm:alfa=1.2} $\alpha=1.2$.
    \subref{app:fig:qsm:fluct:alfa=1},\subref{app:fig:qsm:fluct:alfa=1.2} Average and maximal eigenstate-to-eigenstate fluctuations in Eq.~\eqref{app:eq:fluct} as a function of Hilbert-space dimension for \subref{app:fig:qsm:fluct:alfa=1} $\alpha=1$ and \subref{app:fig:qsm:fluct:alfa=1.2} $\alpha=1.2$.
    The dashed line denotes the ETH expectation with $D^{-1/2}$.
    }
    \label{app:fig:qsm}
\end{figure}
\section{Quantum sun model}\label{sec:qsm}
Now we consider the QSM [\cref{eq:qsm}], for which the Hilbert space of $L_{\rm tot}$ lattice sites (or $L_{\rm tot}$ spins) is more natural, since it is composed of few-body observables.
\subsection{ETH in the QSM}\label{sec:qsm:eth}
Recent works established that for $\alpha=1$, ETH is satisfied in the QSM~\cite{Kliczkowski2024,Swietek2024,Swietek2025a}.
However, a similar analysis for $\alpha>1$ has not yet been carried out.
In this section, we study the diagonal  matrix elements of the operator $\hat{S}^z_{q(L)}\hat{S}^z_L$ for $\alpha>1$ in the Hamiltonian eigenbasis. $q(L)$ is the index of a randomly chosen spin in the ergodic grain of the QSM.
In \cref{app:fig:qsm:alfa=1,app:fig:qsm:alfa=1.2}, we 
show the diagonal matrix elements of this observable at $\alpha=1$ and $\alpha=1.2$, respectively.
These results suggest that the diagonal part of  ETH holds in the ergodic regime for $\alpha\geq1$.
In \cref{app:fig:qsm:fluct:alfa=1,app:fig:qsm:fluct:alfa=1.2}, we study the eigenstate-to-eigenstate fluctuations
\begin{equation}\label{app:eq:fluct}
    z_n = |O_{n+1,n+1} - O_{n,n}|\,,
\end{equation}
where $O_{n,n}=\mel{E_n}{\qmoperator O}{E_n}$, with $\ket{E_n}$ being the Hamiltonian eigenstates sorted by their eigenenergy $E_1 < \ldots < E_n < E_{n+1} <\ldots < E_D$.
Specifically, we consider the average eigenstate-to-eigenstate fluctuations $z_\mathrm{av}=\langle z_n \rangle$ and the maximal outlier $z_\mathrm{max} = \max_n z_n$, where $\langle\ldots\rangle$ and $\max_n$ denote the average over Hamiltonian eigenstates and the maximal value extracted from the central $50\,\%$ of the spectrum, respectively.
Even without disorder averaging, the data clearly indicate that the average fluctuations and the maximal outliers decay with $D^{-1/2}$, as expected from ETH~\cite{Kim2014}.
In order to study ETH in full detail, one needs to consider a set of distinct observables, which is beyond the scope of this work.
Based on our results, we conjecture that observables that are diagonal in the basis in which we express our model as a graph conform with ETH.

\subsection{wEBT in the QSM}\label{sec:qsm:webt}
\begin{figure}[t]
    \centering
    {\phantomsubcaption\label{fig:qsm_typ_gec_tau:a}}
    {\phantomsubcaption\label{fig:qsm_typ_gec_tau:b}}
    \includegraphics[width=\columnwidth]{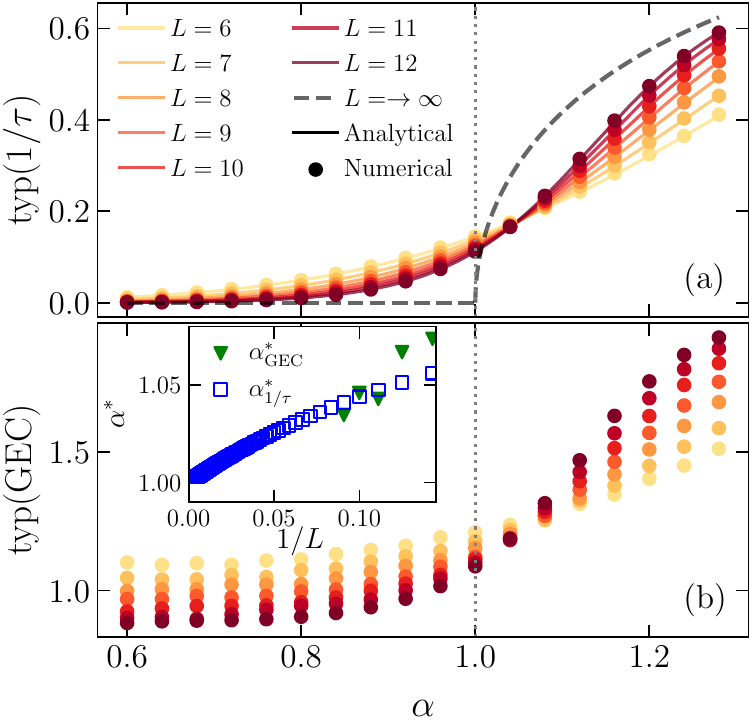}
    \caption{
        Quantum sun model. \subref{fig:qsm_typ_gec_tau:a}  Typical value of the coefficients $1/\tau$ in the short-time expansion of the density autocorrelation functions obtained numerically (markers) and analytically (solid lines).
        The dashed curve is the prediction in the thermodynamic limit ($L\to \infty$), see \cref{app:eq:qsm:typical_tau:alfa_less0,app:eq:qsm:typical_tau:alfa_more0} in \cref{app:tau:qsm}.
        \subref{fig:qsm_typ_gec_tau:b} Typical values of the $\GEC$ distribution for different values of $\alpha$ and different system sizes calculated numerically.
        The vertical dotted line in both panels denote the wEBT.
        In the inset, we show the crossing points extracted from $\mathrm{typ}(1/\tau)$ and $\mathrm{typ}(\GEC)$ between two consecutive system sizes $L$ and $L+1$ as a function of system size.
        The crossing points extracted from $\mathrm{typ}(1/\tau)$ extrapolate to $\alpha=1$ in the thermodynamic limit.
    }
    \label{fig:qsm_typ_gec_tau}
\end{figure}

For the QSM, we consider the density autocorrelation function on the site that is the weakest coupled to the dot, $\ell=L$, where the density operator is defined as $\hat{n}_\ell=\hat{S}^z_\ell+\frac{1}{2}$.
The $\GEC$ is calculated in the basis of spin product states that are joint eigenstates of all $\qmoperator S^z_{\ell}$ and $\qmoperator S^z_{k(\ell)}$.
Note that in the calculations of the moments of the $\GEC$, we assume a very small displacement of the outer spins, $\zeta\ll1$, which does not change the physics of the model.
We compute the entire distribution of both $1/\tau$ and $\GEC$ numerically on small systems and show the typical values of both distributions in \cref{fig:qsm_typ_gec_tau:a} and \cref{fig:qsm_typ_gec_tau:b}, respectively.
For the typical value of $1/\tau$, we also display the corresponding analytical predictions (see \cref{app:tau:qsm} for the derivation).

For $\alpha \leq 1 $, $\mathrm{typ}(1/\tau)$ decreases with system size and increases for $\alpha > 1$.
We find that $\mathrm{typ}(1/\tau)$ shows a crossing point for finite system sizes, drifting towards $\alpha = 1$ when $L\to \infty$ [see inset of \cref{fig:qsm_typ_gec_tau:b}], in contrast to the RPM, where the crossing point is independent of matrix dimension.
An analogous behavior can be seen in the mean and variance of the $\GEC$ as illustrated in~\cref{app:qsm:gec:variance}.

For $L\to\infty$, the typical value changes from $\mathrm{typ}(1/\tau)=0$, for $\alpha\leq1$, to a finite value (see \cref{app:tau:qsm} for details), given by
\begin{equation}\label{eq:qsm:typical_tau}
    \left.\lim_{L\to\infty}
    \mathrm{typ}\qty(\frac{1}{\tau})
    \right|_{\alpha>1}
    =
    \sqrt{1 - \alpha^{-2}}+O(\zeta^2)
    \,.
\end{equation}
above $\alpha=1$.
This result is presented using black dashed lines in \cref{fig:qsm_typ_gec_tau:a}.
We note that the behavior of $\mathrm{typ}(1/\tau)$  in the thermodynamic limit across the wEBT is different for the QSM compared to the RPM.
In the former, we observe a continuous variation of $\mathrm{typ}(1/\tau)$ at the wEBT, yet a discontinuity in its derivatives, while the latter is characterized by a discontinuous jump of typical value of $1/\tau$ at the wEBT.

Lastly, we study the typical value of the $\GEC$ for the same system sizes and range of $\alpha$ values in \cref{fig:qsm_typ_gec_tau:b}.
In contrast to $\mathrm{typ}(1/\tau)$, we have not found an analytical calculation of  $\mathrm{typ}(\GEC)$.
In the inset of \cref{fig:qsm_typ_gec_tau:b}, we show the crossing points extracted from $\mathrm{typ}(1/\tau)$ (using the analytical prediction) and from $\mathrm{typ}(\GEC)$ (obtained numerically).
We find that both crossing points drift to smaller values of $\alpha$.
The crossing point for $\mathrm{typ}(1/\tau)$ converges to the wEBT at $\alpha=1$, while the crossing point of $\mathrm{typ}(\GEC)$ still shows larger finite-size effects for the feasible system sizes than the one from $\mathrm{typ}(1/\tau)$. 
Nevertheless, we expect that for increasing system sizes, the $\mathrm{typ}(\GEC)$ will move to the actual wEBT at $\alpha=1$. Already on small systems, these crossing points are
at $\alpha\approx1.04$.

Since one can compute other moments of $p(\GEC)$ on large systems, 
in \cref{app:qsm:gec:variance}, we study the mean and the variance of the $\GEC$ distribution.
We observe identical crossing points on fixed system sizes that flow towards the wEBT at $\alpha=1$. In the thermodynamic limit, both the mean and variance of the $\GEC$ distribution exhibit a discontinuous jump at the transition.
This also applies to the RPM (results not shown here).

The QSM, similar to the RPM, does not have a local diffusive mode due to the long-range interactions and the all-to-all coupling of the ergodic grain.
Therefore, the relevant processes that determine the dynamics can be captured at short times and they persist until long times.
This, however, might not be the case for spins closer to the grain, for which the dynamics becomes faster for $\alpha<1$ and paradoxically slower for $\alpha>1$, see \cref{app:tau:qsm}.
A full characterization of this possible site-potential hierarchy is outside the scope of this work.

A similar behavior of the $\GEC$, namely that the moments of $\GEC$ show crossing points that extrapolate to a finite value of a model parameter, is also observed in the quantum East model~\cite{Menzler2025} and the triangular lattice-gas model~\cite{Lan2018}, which is considered in an ongoing work.
However, for the these kinetically constrained models, the behavior of the typical values of $1/\tau$ is more involved than for the models studied in this work, leaving open the exact interpretation of the signatures in the GEC in these cases.

\section{Conclusion and Outlook}\label{sec:conclusions}
In this work, we demonstrated that wEBTs can be captured in the short-time expansion of density autocorrelation functions.
Furthermore, we argued that graph-theoretical measures, here the $\GEC$, show similar behavior in the moments of their distribution.
Both the $\GEC$ and the decay rate $1/\tau$ are non-eigenstate measures and, for the models investigated here, their moments can be calculated analytically on large finite systems and in some cases, even in the thermodynamic limit.
Exploiting this possibility, we showed that the typical value of $1/\tau$ quantitatively captures known wEBTs in the RPM and the QSM. Moreover, we argued that the behavior of the GEC is inherited
from that of the decay rates and therefore, the  
typical value of  the $\GEC$ captures the wEBT as well.
To summarize, the moments of the $1/\tau$ are, for the type of wEBTs studied here, extremely useful to identify points where ETH is weakly violated as a system parameter is varied.

Another open question concerns a potential classification of wEBTs beyond those that occur due to fading ergodicity. 
An obvious  candidate class for wEBTs are quantum spin-glass transitions. 
In our ongoing studies of kinetically constrained models, we find crossing points of the moments of the $\GEC$, while the connection to  the short-time expansion is not as straightforward to establish, leaving this for future work.

The  question of whether our approach can contribute to the discussion of the MBL phase and a proposed prethermal phase~\cite{Long2023} is very interesting as well. 
While we find crossing points in the typical value of the $\GEC$ for the disordered XXZ model (results not shown here), there is no direct connection to the slow-down of autocorrelation functions at short times.
This is not surprising: The quantum sun model was constructed to be an effective model of avalanches in disordered systems, yet in microscopic models, higher-order processes are relevant (see e.g., ~\cite{Ohgane2026}).
These questions  are left for future work.

We finish by emphasizing the methodological strength of our analysis. Since the typical value of the $1/\tau$ and GEC can be obtained efficiently for asymptotically large system sizes, one can study regimes with strong finite-size corrections, beyond the reach of exact diagonalization or even matrix-product state methods~\cite{Schollwoeck2011}.
This renders our approach of high potential use for detecting weak ergodicity-breaking transitions in two and higher dimensions,
in particular, since exact-diagonalization studies of higher-dimensional systems are severely limited for interacting systems.

\acknowledgements 
We thank S. Kehrein, M. Mierzejewski, K. Ohgane, F. Pollmann,  and L. Vidmar for useful discussions.
This work was funded by the Deutsche Forschungsgemeinschaft (DFG, German Research Foundation) -- 499180199, 436382789, and 493420525; via FOR 5522 and large-equipment grants (GOEGrid cluster) and Germany's Excellence Strategy -- EXC-2111 -- 390814868.
We acknowledge the usage of Claude Sonnet 5 (Anthropic), provided by the resource of the GWDG, in calculating the typical value of the $\GEC$ and providing a shorter derivation for the mean and the variance of the $\GEC$ for the Rosenzweig-Porter model.
All results produced by AI have been independently checked by the authors.

Research data associated with this article is available on Zenodo~\cite{this_zenodo}.

\appendix

\section{Properties of $\chi^2_k$ distributions}\label{app:chi2}
The cornerstone of the analytical derivations of the moments of the $1/\tau$ and $\GEC$ distributions lies in identifying that those quantities can be represented as chi-squared variables with $k$ degrees of freedom or sums of such variables, respectively.
In the following, we will need several ingredients to derive their moments:
\begin{enumerate}
    \item Given a random variable $X$ we define its typical value as $\mathrm{typ}(X) = \exp{ \mathbb{E}[\ln{X}] }$. Then, the typical value of $aX$, with $a\in\mathbb{R}_+$, is directly given as $\mathrm{typ}(aX) = a\cdot\mathrm{typ}(X)$.
    \item Let $X$ be a random variable, which is $\chi^2_k$ distributed. The relevant moments of $X$ are $\mathbb{E}[X]=k$, $\mathrm{var}(X)=2k$ and $\mathrm{typ}(X)=2e^{\psi(k/2)}$, where $\psi(x)$ is the digamma function.
    \item The $\chi^2_k$ distribution is identical to the Gamma distribution with shape parameter $k/2$, \ie $\chi_k^2\equiv \Gamma(\frac{k}{2},2)\equiv2\Gamma(\frac{k}{2},1)$
    \item Let $X_1\sim\Gamma(k_1,1)$ and $X_2\sim\Gamma(k_2,1)$ be independent random variables.
    Define $X=X_1+X_2$ and $B = X_1 / X$.
    Then, using the properties of the Beta-Gamma algebra, $X$ and $B$ are exactly independent and follow a $\Gamma(k_1+k_2,1)$ and $\mathcal{B}{(k_1,k_2)}$ (beta) distribution, respectively~\cite{Dufresne1998}.
\end{enumerate}

\section{Calculations of moments of $1/\tau$ and $\GEC$}\label{app:gec}
In this section, we sketch the analytical approach to derive the moments of the $1/\tau$ and $\GEC$ defined in \cref{eq:tau:def,eq:gec:def}, respectively.
As outlined in the main text, we rescale the Hamiltonian as $\qmoperator H \to (\qmoperator H - \mu\mathbb{1})/\sigma$, with $\mu = \Tr(\qmoperator H)/D$ and $\sigma^2 = \Tr(\qmoperator H^2)/D-\mu^2$.
For a single system, the denominator in \cref{eq:gec:def} takes care of normalizing the Hamiltonian to unit variance, while the traceless condition has to be imposed manually.
For ensembles of Hamiltonians (as, e.g., in the case of the RPM), we impose these conditions \emph{on average}, meaning $\qmoperator H \to (\qmoperator H - \overline{\mu} \mathbb{1})/\sqrt{\overline{ \sigma^2 }}$, where $\overline{\vphantom{\sum}\ldots}$ denotes the average over different Hamiltonian realizations.

For calculating distributions and moments of $1/\tau$ following the above discussion we rewrite \cref{eq:tau:def} as
\begin{equation}\label{eq:tau2}
\begin{split}
    \frac{1}{\tau_{i}^2}
        &=\frac{1}{ \sigma^2 }n_\ell^{(i)}\sum_j(1-n_\ell^{(j)})|H_{ij}|^2\\
        &=\frac{1}{ \sigma^2 }\sum_{j:\ n_\ell^{(j)}=0}|H_{ij}|^2\,,
\end{split}
\end{equation}
where we use the short form $H_{ij} = \mel{i}{\qmoperator H}{j}$ here and in the remainder of the text.
In the last equality in \cref{eq:tau2}, we are only interested in the non-trivial values of $1/\tau_{i}$.
Hence, we discard the prefactor $n_\ell^{(i)}$ and consider the $D/2$ states $\ket{i}$ that are occupied at site $\ell$, while the second sum in \cref{eq:tau2} is limited to states that are unoccupied at site $\ell$, therefore $j\neq i.$

Next, let us rewrite the definition of the $\GEC$ in \cref{eq:gec:def} by employing the matrix elements of the Hamiltonian as
\begin{equation}\label{app:eq:gec:def}
    \GEC(\ket{i}) = \frac{2(\qmoperator H^2)_{ii}-H_{ii}^2}{\frac{1}{D}\Tr{\qmoperator H^2}}\,.
\end{equation}

In this work, we are interested in finding the moments of the $1/\tau$ and $\GEC$ distribution, where we consider the average over basis states $\ket{i}$ and an additional average over Hamiltonian realizations.
Within each ensemble realization, we define a set of random variables $a_i$ [for instance, $a_i\equiv \GEC(\ket{i})$] for each basis state $\ket{i}$, and define the average $\mathbb{E}[a] = \overline{(1/D)\sum_{i=1}^Da_i}$, where $\overline{\vphantom{\sum}\ldots}$ denotes an average over Hamiltonian realizations, for example $\mathbb{E}[\GEC]=\overline{(1/D)\sum_{i=1}^D\GEC(\ket{i})}$.
Note that since the denominator in \cref{app:eq:gec:def} is independent of the state $\ket{i}$, the average over basis states affects only the numerator of the $\GEC$.
Additionally, when an average over different Hamiltonian realizations is required, we adopt the approximation, where we perform the average independently for the numerator and the denominator.
We denote the numerator by $x_i\equiv 2(\qmoperator H^2)_{ii}-H_{ii}^2$ and approximate the typical value of the $\GEC$ as
\begin{equation}\label{app:eq:gec:approx:typical}
    \mathrm{typ}(\GEC) 
    =
    e^{\mathbb{E}[\ln\GEC]}
    \approx
    \frac{1}{\overline{\sigma^2}}\exp{ \frac{1}{D}\sum_{i=1}^D\overline{\ln{x_i}} }\,.
\end{equation}

To illustrate the approach outlined above, we first consider a Hamiltonian matrix $\hat H=\hat{M}$ drawn from the GOE.
Therefore, the moments of the matrix elements are $\overline{M_{ij}}=0$, $\overline{M_{ii}^2}=2$ and $\overline{M_{ij}^2}=1$ for $i\neq j$.
The width of a GOE matrix is known and given by $\frac{1}{D}\overline{\Tr{\qmoperator M^2}}=D+1$~\cite{Mehta1991}.

The distribution of $1/\tau$ can be deduced exactly in this case.
The sum over $D/2$ standard Gaussian random numbers in \cref{eq:tau2} follows a $\chi^2_{D/2}$ distribution.
This allows for a direct evaluation of the moments of $1/\tau$ from the properties of a $\chi^2_{D/2}$ distribution.
When substituting the variance of the spectrum in \cref{eq:tau2} with the averaged expression for the GOE matrix, i.e., $\sigma^2\rightarrow\overline{\sigma^2}=D+1$, the typical value of $1/\tau$ can be expressed as
\begin{equation}\label{eq:tau:goe:typical}
    \mathrm{typ}\qty(\frac{1}{\tau})=\sqrt{\frac{2e^{\psi(D/4)}}{D+1}}=\frac{1}{\sqrt{2}}-\frac{3}{2\sqrt{2}D}+O\left(\frac{1}{D^2}\right)\,,
\end{equation}
where we applied a large-$D$ expansion and $\psi(x)$ is the digamma function.
Here and in all subsequent derivations we will use the fact that $\mathrm{typ}\qty(1/\tau)=\sqrt{\mathrm{typ}\qty(1/\tau^2)}$.

The numerator of the $\GEC$ can be also computed exactly for a GOE matrix.
The numerator of the $\GEC$ is $x_i=M_{ii}^2+2\sum_{j\neq i}M_{ij}^2$, which is $\chi^2_D$ distributed.
As a consequence, the typical value of the $\GEC$ is 
\begin{equation}\label{eq:gec:goe:typical}
    \mathrm{typ}(\GEC)=\frac{4}{D+1}e^{\psi(D/2)} = 2 - \frac{4}{D} + O\left(\frac{1}{D^2}\right)\,.
\end{equation}
The results in \cref{eq:tau:goe:typical,eq:gec:goe:typical} provide a benchmark for the typical values of the $1/\tau$ and of the $\GEC$ distributions in Hamiltonians sampled from a GOE.
Following a similar approach, we calculate these typical values for the RPM in the next section.

\section{Derivation of typical values of $1/\tau$ and $\GEC$ in the RPM}\label{app:gec:rpm}
It has been shown that the width of the spectrum of the RPM defined in \cref{eq:rpm:ham}   (whose mean is $\mu=0$) is~\cite{Venturelli2023}
\begin{equation}\label{app:eq:rp:var}
    \overline{\sigma^2}=\frac{1}{D}\Tr{\qmoperator H^2_\mathrm{RPM}} = 1+D^{1-\gamma}+D^{-\gamma}\,.
\end{equation}
Let us note that the following calculations for the RPM do not depend on the particular form of the basis states $\ket{i}$ as we only take into account the moments of the elements of the matrices $\hat{H}_0$ and $\hat{M}$.

Finding the distribution of $1/\tau$ for the RPM can be directly derived from the distribution of its matrix elements.
The short-time coefficients in \cref{eq:tau2} are (already making use of the substitution $\sigma^2\rightarrow\overline{\sigma^2}$)
\begin{equation}\label{eq:tau:RPM}
    \frac{1}{\tau_i^2}
    =\frac{1}{\overline{\sigma^2}}D^{-\gamma}\sum_{j\neq i:\ n_\ell^{(j)}=0}|M_{ij}|^2 \,.
\end{equation}
The sum in \cref{eq:tau:RPM} is a random varable, which follows a $\chi^2_{D/2}$ distribution, as in the case of a GOE matrix.
Following again the properties of a $\chi^2_k$ distribution outlined in \cref{app:chi2}, we arrive at a formula for the typical value of $1/\tau$ for the RPM:
\begin{equation}
    \mathrm{typ}\qty(\frac{1}{\tau}) = \sqrt{\frac{2D^{-\gamma}e^{\psi(D/4)}}{1 + D^{1-\gamma} + D^{-\gamma}}}\,.
\end{equation}

Next, we consider the $\GEC$ distribution.
The numerator of the $\GEC$ for the RPM can be expressed as
\begin{equation}\label{app:eq:rp:x_i}
    x_i = \left(\epsilon_i + D^{-\gamma/2}M_{ii} \right)^2
    + 2D^{-\gamma}\sum_{j\neq i}M_{ij}^2\,.
\end{equation}
Let us denote the individual terms in \cref{app:eq:rp:x_i} as
\begin{align}
    u_i &= \left(\epsilon_i + D^{-\gamma/2}M_{ii}\right)^2\\
    v_i &= 2D^{-\gamma}\sum_{j\neq i}M_{ij}^2\,.
\end{align}
As a (sum of) squared Gaussian variables, $u_i$ and $v_i$ follow a rescaled $\chi^2_k$ distribution.
Taking into account the discussion in \cref{app:chi2}, we introduce the scaled variables $\tilde{u}_i\equiv u_i/(2\sigma_u^2)$, where $\sigma_u^2 = 1 + 2D^{-\gamma}$, and $\tilde{v}_i\equiv v_i/(4D^{-\gamma})$, which are $\Gamma\qty(\frac{1}{2},1)$ and $\Gamma\qty(\frac{D-1}{2},1)$ distributed, respectively.

The calculation of the typical value of the $\GEC$ requires the Beta-Gamma algebra for two $\Gamma(k,1)$-distributed variables, see \cref{app:chi2}.
We can rewrite the numerator of the $\GEC$ by defining $T_i = \tilde{u}_i+\tilde{v}_i$, which is $\Gamma\qty(\frac{D}{2},1)$ distributed, and the variable $F_i=\tilde{u}_i/T_i$, which follows a beta distribution $\mathcal{B}{\qty(\frac{1}{2},\frac{D-1}{2})}$, as
\begin{equation}
\begin{split}
       x_i &= u_i + v_i = 2\sigma_u^2\tilde{u}_i + 4D^{-\gamma}\tilde{v}_i\\
&= 2\tilde{u}_i + 4D^{-\gamma}\qty[\tilde{u}_i + \tilde{v}_i]\\
&= 2T_i F_i + 4D^{-\gamma}T_i = 2T_i\qty(F_i + 2D^{-\gamma})\,.
\end{split}
\end{equation}
Therefore, the expectation value of the logarithm of $x_i$ is
\begin{equation}
    \mathbb{E}[\ln{x}]
    =
    \mathbb{E}[\ln(2T)]
    +
    \underbrace{\mathbb{E}[\ln\qty(F+2D^{-\gamma})]}_{J(\gamma,D)}\,.
\end{equation}
The first expectation value can be trivially computed, since $2T_i$ follows a $\chi^2_D$ distribution that yields $\mathbb{E}[\ln(2T)]=\ln{2}+\psi(D/2)$.
The second term can be expressed in integral form using the probability distribution of the beta ensemble
\begin{equation}
    f_F(z) = \frac{1}{A}\frac{(1-z)^{\frac{D-1}{2}-1}}{\sqrt{z}}\,,
\end{equation}
where $z\in(0,1)$ and $A=B\qty(\frac{1}{2},\frac{D-1}{2})$ is the normalization constant and $B\qty(a,b)$ is the Beta function.
Hence, the last term is
\begin{equation}\label{eq:J_kappa}
    \begin{split}
        J(\gamma,D) = \frac{1}{A}\int_0^1 dz \frac{(1-z)^{\frac{D-1}{2}-1}}{\sqrt{z}}\ln\qty(z+2D^{-\gamma})\,.
    \end{split}
\end{equation}
As a result, the typical value of the $\GEC$ is
\begin{equation}
    \mathrm{typ}(\GEC) 
    = \frac{2e^{\psi(D/2)}e^{J(\gamma,D)}}{1+D^{-\gamma}(D+1)}\,,
\end{equation}
where we replaced $\sigma^2\to\overline{\sigma^2}$ in the definition of the $\GEC$ as in \cref{app:eq:gec:approx:typical}.
The integral $J(\gamma,D)$ does not have a closed analytical expression and it has to be computed numerically.
Due to numerical instabilities, this approach allows us to exactly compute numerically system sizes up to only $D=2^{24}$.
In order to reach larger system sizes, we make a few approximations.
For large $D\gg1$ the beta distribution is asymptotically $\frac{D}{2}\mathcal{B}{\qty(\frac{1}{2},\frac{D-1}{2})}\to\Gamma\qty(\frac{1}{2},1)\equiv\frac{1}{2}\chi_1^2$ with relative corrections $\mathcal{O}(1/D)$.
In this limit, $F/D$ is approximately $\chi^2_1$ distributed, which we approximate by $Z^2$ in \cref{eq:J_kappa}, where $Z\sim N(0,1)$.
We define the function $g(\Lambda)=\mathbb{E}\qty[\ln\qty(Z^2+\Lambda)]$ with $\Lambda = 2D^{1-\gamma}$, and
using the expansion of the digamma function $\psi(D/2)=\ln({D/2})+O(1/D)$, we find a simpler expression for the typical value of the $\GEC$
\begin{equation}
    \mathrm{typ}(\GEC) \simeq \frac{e^{g(\Lambda)}}{1+D^{-\gamma}(D+1)}\,.
\end{equation}
Next, we use the Frullani integral form of $\ln{x}$
\begin{equation}
    \ln\qty[Z^2+\Lambda]=\int_0^\infty\frac{dt}{t}\qty[e^{-t}-e^{-\Lambda t}e^{-Z^2t}]\,.
\end{equation}
Since $Z$ is a Gaussian random variable, $\mathbb{E}[e^{-Z^2t}]=\qty(1+2t)^{-1/2}$, and exchanging the integral with the expectation value (allowed by Fubini's theorem), we arrive at the integral representation of $g(\Lambda)$
\begin{equation}
    g(\Lambda) = \int_0^\infty \frac{dt}{t}\qty(e^{-t}-\frac{e^{-\Lambda t}}{\sqrt{1+2t}})\,.
\end{equation}
This approach allows us to estimate the typical value of the $\GEC$ for very large systems with a marginally small correction $\mathcal{O}(1/D)$ without the need of exact diagonalization.

Now, we are also able to find the asymptotic behavior of $\mathrm{typ}(\GEC)$ in the thermodynamic limit.
Using the scaling variable $\Lambda=2D^{1-\gamma}$, we identify two distinct regimes:
\begin{enumerate}[label=(\alph*)]
    \item When $\gamma<1$, the scaling variable $\Lambda$ diverges, \ie $\Lambda\to\infty$ for $D\to\infty$, such that we can expand the logarithm $\ln(Z^2+\Lambda)\approx\ln{\Lambda} + O(\Lambda^{-1})$, hence $g(\Lambda)\simeq\ln\Lambda=\ln\qty(2D^{1-\gamma})$.
    The variance $\overline{\sigma^2}$ asymptotically scales as $\overline{\sigma^2}\sim D^{1-\gamma}$.
    This leads to the typical value
    \begin{equation}\label{eq:rpm:gec:typical:gamma_less1}
        \left.\lim_{D\to\infty}\mathrm{typ}(\GEC)\right|_{\gamma<1} = 2\,.
    \end{equation}
    \item 
    Above the wEBT for $\gamma>1$, the scaling variable $\Lambda$ vanishes, \ie $\Lambda\xrightarrow[]{D\to\infty}0$.
    In this case, we find $g(0)=-c_E-\ln{2}$, where $c_E\approx0.57721$ is Euler's constant.
    The variance for $D\to\infty$ in this regime becomes $\overline{\sigma^2}\to1$, thus the typical value of $\GEC$ converges to
    \begin{equation}\label{eq:rpm:gec:typical:gamma_more1}
        \left.\lim_{D\to\infty}\mathrm{typ}(\GEC)\right|_{\gamma>1} = \frac{e^{-c_E}}{2}\,.
    \end{equation}
\end{enumerate}
The results in \cref{eq:rpm:gec:typical:gamma_less1,eq:rpm:gec:typical:gamma_more1} show the discontinuity of the typical value of $\GEC$ at the wEBT in the RPM, see \cref{fig:RP_var_gec_tau:b}.

\section{Derivation of typical value of $1/\tau$ in the QSM}\label{app:tau:qsm}
To find properties of the distribution of $1/\tau$ we use the definition in \cref{eq:tau2} in terms of the commutator as (where we choose $\ell=L$)
\begin{equation}
    \frac{1}{\tau_{i}^2} = -\frac{1}{\overline{{\sigma^2}}}\mel{i}{[\hat{H},\hat{n}_L]^2}{i}\,,
\end{equation}
where we again substituted ${\sigma^2}\to\overline{\sigma^2}$.
This commutator can be computed exactly for the QSM.
The only term not commuting with a local density operator on $L$, which is outside the grain, is the spin-flip term, namely
\begin{equation}
    [\hat{H},\hat{n}_L] = g_0\alpha^{u_L}[\hat{S}^x_{k(L)}\hat{S}^x_L,\hat{n}_L]
    =
    g_0\alpha^{u_L}\hat{S}^x_{k(L)}[\hat{S}^x_L,\hat{n}_L]\,,
\end{equation}
where we used the fact that $\hat{S}^x_{k(L)}$ commutes with all density operators on sites that are outside the grain.
Using the algebra of Pauli matrices, we directly find that the commutator reduces to
\begin{equation}
    [\hat{H},\hat{n}_L] =
    ig_0\alpha^{u_L}\hat{S}^x_{k(L)}\hat{S}^y_L\,.
\end{equation}
Therefore, the short-time coefficient becomes initial-state independent with a value
\begin{equation}
    \frac{1}{\tau_{i}^2} = \frac{g_0^2}{16\overline{\sigma^2}}\alpha^{2u_L}
    \quad\to\quad
    \frac{1}{\tau_{i}} = \frac{g_0}{4\sqrt{\overline{\sigma^2}}}\alpha^{u_L}
    \,.
\end{equation}
We used the fact that the spin operators satisfy $(\hat{S}^\kappa_L)^2=\frac{1}{4}$ for $\kappa=x,y,z$.
As a result, the distribution of $1/\tau$ for the QSM is solely given through the distributions of $u_L$.
This leads to an analytical estimate of the typical value of $1/\tau$ of the form
\begin{equation}\label{eq:qsm:tau}
	\mathrm{typ}\qty(\frac{1}{\tau}) = \frac{g_0}{4\sqrt{\overline{\sigma^2}}}\alpha^{L-1}\,.
\end{equation}
Interestingly, an identical derivation for different spins $\ell$ leads to a hierarchy of relaxation rates with the typical value being $\mathrm{typ}(1/\tau)\propto\alpha^{\ell-1}$.
Spins that are closer to the grain might not capture the wEBT in the same way.

Next, we need to find $\overline{\sigma^2}$.
The Hilbert space of the QSM can be decomposed into the degrees of freedom of the thermal grain and the localized spins $\mathcal{H}=\mathcal{H}_\mathrm{grain}\otimes\mathcal{H}_\mathrm{loc}$.
Hence, any product state in the full Hilbert space can be viewed as a tensor product of the associated subspaces $\ket{i}=\ket{n}\otimes\ket{\mu}$, where $\ket{n}\in\mathcal{H}_\mathrm{grain}$ and $\ket{\mu}\in\mathcal{H}_\mathrm{loc}$.
With this decomposition, we can express the terms of the square of the Hamiltonian in \cref{eq:qsm} as
\begin{equation}\label{eq:qsm:H2}
\begin{split}
    (\qmoperator H^2)_{ii} = &
    (\qmoperator R^2)_{nn} + \frac{g_0^2}{16}\sum_{\ell}\alpha^{2u_\ell}\\
   &+\sum_{\ell,\ell'}h_\ell h_{\ell'}\mel{\mu}{\qmoperator S^z_\ell \qmoperator S^z_{\ell'}}{\mu}\\
   & + 2R_{nn}\sum_\ell h_\ell\mel{\mu}{\qmoperator S^z_\ell}{\mu}\,.
\end{split}
\end{equation}
Using the identity for spin operators, $(1/D)\Tr(\qmoperator S^z_\ell \qmoperator S^z_{\ell'})=\frac{1}{4}\delta_{\ell\ell'}$, and their tracelesness, $\Tr(\qmoperator S^z_\ell)=0$, we find the width of the spectrum for a single realisation as
\begin{equation}\label{app:eq:qsun:sigma}
    \sigma^2 = 1 + \frac{g_0^2}{16}\sum_{\ell}\alpha^{2u_\ell}
    + \frac{1}{4}\sum_\ell h_\ell^2\,.
\end{equation}
We note that the Hamiltonian in \cref{eq:qsm} is not exactly traceless due to the small size of the random grain.
However, on average, the contribution from the trace in the last term of \cref{eq:qsm:H2} vanishes.
We also used the fact that the operator $\qmoperator R$ is a normalized GOE matrix.

Let us now find the 
ensemble-averaged variance over energies.
In the second term of \cref{app:eq:qsun:sigma}, $u_\ell$ is a uniform random variable, hence we can compute the average (and any higher moments) of $\alpha^{ku_\ell}$, resulting in ($\forall \ell>1$)
\begin{equation}\label{app:eq:qsun:u_l:moments}
\begin{split}
    \overline{\alpha^{ku_\ell}} &= \int_{\ell-1-\zeta}^{\ell-1+\zeta}\frac{dx}{2\zeta} \alpha^{kx}
    = \frac{\alpha^{k\zeta}-\alpha^{-k\zeta}}{2k\zeta\ln(\alpha)}\alpha^{k(\ell-1)}\\
  &= \frac{\sinh(k\zeta\ln(\alpha))}{k\zeta\ln(\alpha)}\alpha^{k(\ell-1)}
    \equiv G(k\zeta,\alpha)\alpha^{k(\ell-1)}\,.
\end{split}
\end{equation}
Here, we define the function $G(k\zeta,\alpha)=\mathrm{sinhc}(k\zeta\ln(\alpha))$, which has the limiting value $\lim_{\zeta\to0}G(k\zeta,\alpha)=1$.
Then, the sum over $\ell$ can be trivially computed using the properties of a geometric series with ratio $\alpha^2$ (we also need to exclude $u_1=0$ from the average):
\begin{equation}
    \begin{split}
    \sum_{\ell=1}^L\overline{\alpha^{2u_\ell}}
    &= 1 + \sum_{\ell=2}^L\overline{\alpha^{2u_\ell}}
    = 1 + G(2\zeta,\alpha)\sum_{\ell=2}^L \alpha^{2\ell-2} \\
    &= 1 + G(2\zeta,\alpha)\sum_{\ell'=1}^{L-1} \qty(\alpha^2)^{\ell'}\\
    &= 1+G(2\zeta,\alpha)\qty(\frac{1-\alpha^{2L}}{1-\alpha^2}-1)\,.
    \end{split}
\end{equation}
This leads to the average energy variance
\begin{equation}\label{app:eq:qsm:sigma:averaged}
\begin{split}
	\overline{\sigma^2} =& 1 + \frac{g_0^2}{16}\qty[1+G(2\zeta,\alpha)\qty(\frac{1-\alpha^{2L}}{1-\alpha^2}-1)]\\
	&+\frac{L}{4}\qty(h^2+\frac{1}{3}W^2)\,,
\end{split}
\end{equation}
where we used the results known for a uniform distribution $\overline{h_\ell^2}=h^2+\frac{1}{3}W^2$.
Plugging this expression into \cref{eq:qsm:tau} leads to the final formula for the typical value of $1/\tau$.

We can estimate the behavior of the typical value in the thermodynamic limit.
Specifically, for $\alpha\leq1$, it follows trivially that
\begin{equation}\label{app:eq:qsm:typical_tau:alfa_less0}
    \left.\lim_{L\to\infty}
    \mathrm{typ}\qty(\frac{1}{\tau})
    \right|_{\alpha\leq1}=0\,,
\end{equation}
since $\sigma^2\propto L$, while $\alpha^{L-1}\ll1$.
On the ergodic side, for $\alpha>1$, the typical value simplifies to
\begin{equation}\label{app:eq:qsm:typical_tau:alfa_more0}
    \left.\lim_{L\to\infty}
    \mathrm{typ}\qty(\frac{1}{\tau})
    \right|_{\alpha>1}
    =
    \sqrt{\frac{1 - \alpha^{-2}}{G(2\zeta,\alpha)}}
    \,.
\end{equation}
In our simulation we choose $\zeta=0.2$, hence, we can approximate to high accuracy $G(2\zeta,\alpha)\approx1$, leading to the following prediction in the thermodynamic limit
\begin{equation}\label{app:eq:qsm:typical_tau:alfa_more0:small_zeta}
    \left.\lim_{L\to\infty}
    \mathrm{typ}\qty(\frac{1}{\tau})
    \right|_{\alpha>1}
    =
    \sqrt{1 - \alpha^{-2}}+O(\zeta^2)
    \,.
\end{equation}
This result is shown in \cref{fig:qsm_typ_gec_tau:a} using dashed lines.

\section{Variance of the $\GEC$ in the QSM}
\label{app:qsm:gec:variance}
In this section we provide additional results on the mean and the variance of the $\GEC$ for the QSM.
First, we define our averaging strategy for both quantities in the following way:
\begin{equation}\label{app:eq:gec:approx:mean}
    \mathbb{E}[\GEC] \approx \frac{1}{D\overline{\sigma^2}}\overline{\sum_{i=1}^Dx_i}
\end{equation}
and
\begin{equation}\label{app:eq:gec:approx:variance}
    \mathrm{var}(\GEC) \approx \frac{1}{\qty(D\overline{\sigma^2})^2}\overline{\qty(\sum_{i=1}^Dx_i^2
    -
    \qty[\sum_{i=1}^Dx_i]^2)}
    \,.
\end{equation}

\begin{figure}[t]
    \centering
    \includegraphics[width=\columnwidth]{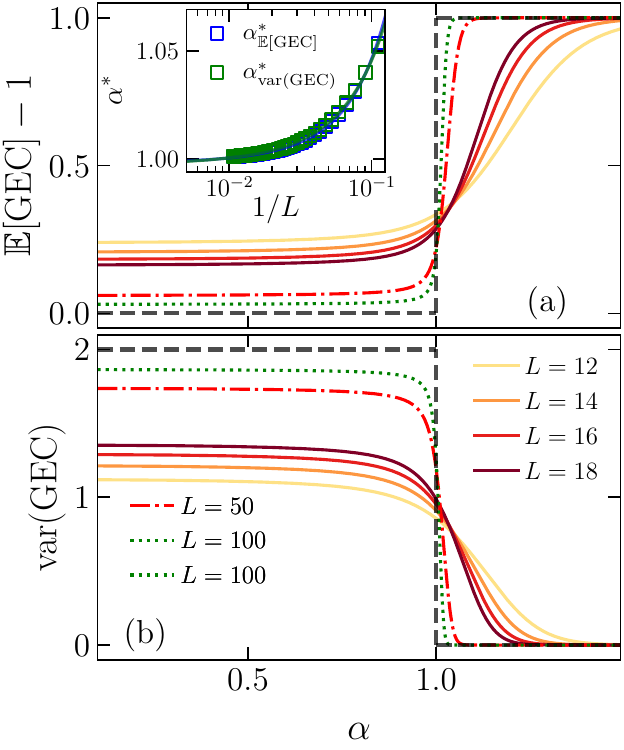}
    {\phantomsubcaption\label{app:fig:qsun:mean}}
    {\phantomsubcaption\label{app:fig:qsun:var}}
    \caption{
    Quantum sun model.
    \subref{app:fig:qsun:mean} Average mean and \subref{app:fig:qsun:var} variance of the $\GEC$ for the QSM for different values of $\alpha$ and system sizes $L$ using the analytical predictions in \cref{app:eq:qsm:GEC:mean2,app:eq:qsm:GEC:var}, respectively.
    We choose the following model parameters $N=3$, $g_0=1$, $\zeta=0.2$, $h=1$ and $W=0.5$.
    The dashed line shows the analytical predictions in the thermodynamic limit $L\to\infty$, see the text for details.
    The inset in \subref{app:fig:qsun:mean} shows crossing points between two system sizes $L$ and $L+2$ for both the mean $\mathbb{E}[\GEC]$ and variance $\mathrm{var}(\GEC)$.
    Already for small system sizes, the crossing points for both quantities nearly coincide and extrapolate to $\alpha\approx0.998$ in the thermodynamic limit.
    }
    \label{app:fig:qsun}
\end{figure}
The calculation of the mean of the $\GEC$ follows the same steps as the computation of $\overline{\sigma^2}$ in \cref{app:tau:qsm}.
The numerator of the $\GEC$ can be readily expressed as
\begin{equation}\label{eq:qsm:x_i}
\begin{split}
    x_i = &
    2(\qmoperator R^2)_{nn} - R_{nn}^2
    + \frac{g_0^2}{8}\sum_{\ell}\alpha^{2u_\ell}\\
   &+\sum_{\ell,\ell'}h_\ell h_{\ell'}\mel{\mu}{\qmoperator S^z_\ell \qmoperator S^z_{\ell'}}{\mu}\\
   & + 2R_{nn}\sum_\ell h_\ell\mel{\mu}{\qmoperator S^z_\ell}{\mu}\,,
\end{split}
\end{equation}
while the mean value of the $\GEC$ is
\begin{equation}
    \label{app:eq:qsm:GEC:mean}
    \mathbb{E}[\GEC] = 2 - \frac{1}{\overline{\sigma^2}}
    \qty[\frac{2}{2^N+1} + \frac{L}{4}\qty(h^2+\frac{1}{3}W^2)]\,,
\end{equation}
where $\overline{\sigma^2}$ is given by \cref{app:eq:qsm:sigma:averaged}.

In this work, we consider only small fluctuations in the position of the outside spins ($\zeta\ll1$), which allows us to approximate $G(k\zeta,\alpha)\approx1$.
While $G(k\zeta, \alpha)$ can be calculated exactly, as above, it is instructive to use this approximation to simplify \cref{app:eq:qsm:GEC:mean}.
In essence, for the discussed systems, where we consider $\zeta = 0.2$, this approximation is highly accurate, with an error of $|1-G(2\zeta,\alpha)|<5\cdot 10^{-3}$ in the considered range of $\alpha$.
This yields the slightly simpler result for the mean of the $\GEC$
\begin{equation}\label{app:eq:qsm:GEC:mean2}
    \mathbb{E}[\GEC] \approx \frac{\frac{2^{N+1}}{2^N+1} + \frac{g_0^2}{8}\frac{1-\alpha^{2L}}{1-\alpha^2}+\frac{L}{4}\qty(h^2+\frac{1}{3}W^2)}{1 + \frac{g_0^2}{16}\frac{1-\alpha^{2L}}{1-\alpha^2}+\frac{L}{4}\qty(h^2+\frac{1}{3}W^2)}
    \equiv
    \mathbb{E}[\GEC]^*
    \,,
\end{equation}
where we define $\mathbb{E}[\GEC]^*$ as the average value of the $\GEC$ under the approximation $\zeta\ll1$.
In \cref{app:fig:qsun:mean}, we show the mean of the $\GEC$ in \cref{app:eq:qsm:GEC:mean2} for different system sizes as a function of $\alpha$.
We observe similar behavior with respect to the typical value the $\GEC$ and the typical value of the decay rates $1/\tau$ studied in the main text.
In the inset of \cref{app:fig:qsun:mean}, we also show the scaling of the crossing points that extrapolate to $\alpha=1$, which is the predicted value of the wEBT.
In contrast to the typical value of $1/\tau$, we find that in the thermodynamic limit the mean $\mathbb{E}[\GEC]$ shows a discontinuous jump from $\mathbb{E}[\GEC]_{\alpha<1}=1$ to $\mathbb{E}[\GEC]_{\alpha>1}=2$.
We expect the same discontinuous jump to occur for $\mathrm{typ}(\GEC)$.

Lastly, let us turn to the calculation of the variance of the $\GEC$ for the QSM.
The square of $x_i$ can be expressed as
\begin{equation}\label{app:eq:qsun:x_i2}
    \begin{split}
        x_i^2 
        =& A_{nn}^2 + \frac{g_0^4}{64}\sum_{\ell,\ell'}\alpha^{2u_\ell}\alpha^{2u_{\ell'}}\\
        &+ \sum_{\ell_1,\ell_2,\ell_3,\ell_4} h_{\ell_1}h_{\ell_2}h_{\ell_3}h_{\ell_4}\mel{\mu}{\qmoperator S^z_{\ell_1}\qmoperator S^z_{\ell_2}\qmoperator S^z_{\ell_3}\qmoperator S^z_{\ell_4}}{\mu}\\
        &+ \qty[4R_{nn}^2 + 2A_{nn} + \frac{g_0^2}{4}\sum_{\ell}\alpha^{2u_\ell}]\sum_{\ell,\ell'}h_\ell h_{\ell'}\mel{\mu}{\qmoperator S^z_{\ell}\qmoperator S^z_{\ell'}}{\mu}\\
        &+\frac{g_0^2}{4}A_{nn}\sum_{\ell}\alpha^{2u_\ell}
        +2A_{nn}R_{nn}\sum_\ell h_\ell\mel{\mu}{\qmoperator S^z_\ell}{\mu}\\
        &+\frac{g_0^2}{2}R_{nn}\sum_{\ell}\alpha^{2u_\ell}\sum_{\ell'} h_{\ell'}\mel{\mu}{\qmoperator S^z_{\ell'}}{\mu}\\
        &+4R_{nn}\sum_{\ell_1,\ell_2,\ell_3} h_{\ell_1}h_{\ell_2}h_{\ell_3}\mel{\mu}{\qmoperator S^z_{\ell_1}\qmoperator S^z_{\ell_2}\qmoperator S^z_{\ell_3}}{\mu}\,,
    \end{split}
\end{equation}
where $A_{nn}=2(R^2)_{nn}-R^2_{nn}$.
Since most of the terms present in \cref{app:eq:qsun:x_i2} have been calculated in the context of the mean of the $\GEC$, we can write (note that the last three terms in \cref{app:eq:qsun:x_i2} vanish in the average since $\mathbb{E}[R_{nn}]=0$)

\begin{equation}\label{app:eq:qsun:x_i2:av:-1}
    \begin{split}
        \mathbb{E}[x^2]^*
        =& \frac{1}{2^N}\sum_n\overline{A_{nn}^2}
        + \frac{g_0^4}{64}\sum_{\ell,\ell'}\overline{\alpha^{2u_\ell}\alpha^{2u_{\ell'}}}\\
        &+ \sum_{\ell_1,\ell_2,\ell_3,\ell_4}\overline{h_{\ell_1}h_{\ell_2}h_{\ell_3}h_{\ell_4}}\frac{1}{2^L}\sum_\mu\mel{\mu}{\qmoperator S^z_{\ell_1}\qmoperator S^z_{\ell_2}\qmoperator S^z_{\ell_3}\qmoperator S^z_{\ell_4}}{\mu}\\
        &+ \qty[4+\frac{4}{2^N+1}+\frac{g_0^2}{4}\frac{1-\alpha^{2L}}{1-\alpha^2}]\frac{L}{4}\qty(h^2+\frac{W^2}{3})\\
        &+\frac{g_0^2}{4}\frac{2^{N+1}}{2^N+1}\frac{1-\alpha^{2L}}{1-\alpha^2}\,,
    \end{split}
\end{equation}
where we used the approximation $G(2\zeta,\alpha)\approx1$ for $\zeta\ll1$.
Let us move to the remaining terms in Eq.~\eqref{app:eq:qsun:x_i2:av:-1}.

The first term can be rewritten as $\overline{A_{nn}^2} = \overline{(R^2)_{nn}^2}-4\overline{(R^2)_{nn}R_{nn}^2}+\overline{R_{nn}^4}$.
Notably, these averages can be computed since all moments of matrix elements from the GOE are known.
Including the normalization of the matrix $\qmoperator R$ and the fourth moment being $\overline{R_{nn}^4} = 3\overline{R_{nn}^2}^2=12/(2^N+1)^2$, we arrive at the expression
\begin{equation}
    \frac{1}{2^N}\sum_n\overline{A_{nn}^2} = \frac{2^{N+2}\qty(2^N+2)}{\qty(2^N+1)^2}\,.
\end{equation}
The second part can be computed using the properties of a geometric series and the result in \cref{app:eq:qsun:u_l:moments}, and we obtain the simple expression
\begin{equation}
    \begin{split}
        \sum_{\ell,\ell'}\overline{\alpha^{2u_\ell}\alpha^{2u_{\ell'}}} 
        &=
        \sum_{\ell}\overline{\alpha^{4u_\ell}}
        +
        \sum_{\ell\neq\ell'}\overline{\alpha^{2u_\ell}}\,\overline{\alpha^{2u_{\ell'}}}\\
        &=\sum_{\ell}\overline{\alpha^{4u_\ell}} + \qty(\sum_{\ell}\overline{\alpha^{2u_\ell}})^2 - \sum_{\ell}\overline{\alpha^{2u_\ell}}^2\\
        &\approx \qty(\frac{1-\alpha^{2L}}{1-\alpha^4})^2\,,
    \end{split}
\end{equation}
where we again used the fact that, for small $\zeta$, $G(2\zeta,\alpha)\approx1$ and $G(4\zeta,\alpha)-F^2(2\zeta,\alpha)\approx0$.
To evaluate the third term, we need to break up the sum over the four indices $\ell_1,\ell_2,\ell_3$ and $\ell_4$ into smaller terms.
Specifically, the only non-zero elements arise when all indices are equal or when indices appear in pairs such as $(\ell_1=\ell_2,\ell_3=\ell_4)$, $(\ell_1=\ell_3,\ell_2=\ell_4)$ and $(\ell_1=\ell_4,\ell_2=\ell_3)$.
Thus, the last term is simplified to
\begin{equation}
\begin{split}
&\sum_{\ell_1,\ell_2,\ell_3,\ell_4}\mathbb{E}\qty[h_{\ell_1}h_{\ell_2}h_{\ell_3}h_{\ell_4}\mel{\mu}{\qmoperator S^z_{\ell_1}\qmoperator S^z_{\ell_2}\qmoperator S^z_{\ell_3}\qmoperator S^z_{\ell_4}}{\mu}]\\
&\quad=
\frac{1}{16}\sum_\ell\mathbb{E}[h_\ell^4]
    +
    \frac{3}{16}\sum_{\ell\neq\ell'}\mathbb{E}[h_\ell^2]\mathbb{E}[h_{\ell'}^2]\,,
\end{split}
\end{equation}
where the factor of three  arises due to the three distinct ways of pairing the indices.
Using known expressions for expectation values in a uniform distribution, we find
\begin{equation}
    \begin{split}
&\sum_{\ell_1,\ell_2,\ell_3,\ell_4}\overline{h_{\ell_1}h_{\ell_2}h_{\ell_3}h_{\ell_4}}\frac{1}{2^L}\sum_\mu\mel{\mu}{\qmoperator S^z_{\ell_1}\qmoperator S^z_{\ell_2}\qmoperator S^z_{\ell_3}\qmoperator S^z_{\ell_4}}{\mu}]\\
    &\quad=
    \frac{L}{16}\qty(h^4+2h^2W^2+\frac{W^4}{5})
    +
    \frac{3L(L-1)}{16}\qty(h^2+\frac{W^2}{3})^2\,.
    \end{split}
\end{equation}
Collecting all terms, we arrive at the second moment of $x_i$ in the approximation $\zeta\ll1$, and using the mean of the $\GEC$ in \cref{app:eq:qsm:GEC:mean2},
we arrive at an expression for the variance of the $\GEC$: 
\begin{equation}\label{app:eq:qsm:GEC:var}
\begin{split}
\mathrm{var}(\GEC)^* =&
\frac{2^{N+3}}{(\overline{\sigma^2})^2(2^N+1)^2} \\
    &+ \frac{L}{16(\overline{\sigma^2})^2}\qty(h^4+2h^2W^2+\frac{W^4}{5})\\
       &+ \frac{2L^2-3L}{16(\overline{\sigma^2})^2}\qty(h^2+\frac{W^2}{3})^2\\
        &+ \frac{2L}{(\overline{\sigma^2})^2(2^N+1)}\qty(h^2 + \frac{W^2}{3})\,.
\end{split}
\end{equation}
It is remarkable that the $\alpha$-dependence is solely given by the denominator.

We study \cref{app:eq:qsm:GEC:var} for different system sizes in \cref{app:fig:qsun:var} as a function of $\alpha$.
Again, there is a crossing point for different system sizes that flows towards the wEBT.
In the thermodynamic limit, the variance $\mathrm{var}(\GEC)$ has a step-function at the wEBT much like the mean $\mathbb{E}[\GEC]$.
In the inset of \cref{app:fig:qsun:mean}, we show the crossing points of both the mean $\mathbb{E}[\GEC]$ and the variance $\mathrm{var}(\GEC)$, which nearly overlap and extrapolate to the wEBT at $\alpha=1$.
For a detailed analysis of the subleading corrections and the error estimates of our averaging strategy in \cref{app:eq:gec:approx:mean,app:eq:gec:approx:variance}, we refer the reader to an  earlier version of the manuscript~\cite{Menzler2026_v2}.
Note further that the mean and variance of the GEC can be computed for the RPM as well, exhibiting a discontinuous behavior at the wEBT~\cite{Menzler2026_v2}. 

\bibliography{references.bib}
\end{document}